\documentclass[10pt,conference]{IEEEtran}

\usepackage{cite}
\usepackage{graphicx}
\usepackage{psfrag}
\usepackage{subfigure}
\usepackage{flushend}
\usepackage{url}
\usepackage{amsmath}
\usepackage{array}
\usepackage{graphics}
\usepackage{epsfig}
\usepackage{amsbsy}
\usepackage{amssymb}
\usepackage{amsthm}

\usepackage{framed}
\usepackage{algorithm}
\IEEEoverridecommandlockouts

\begin{document}

\title{\LARGE Block Outlier Methods for Malicious User Detection in Cooperative Spectrum Sensing}
\author{\IEEEauthorblockN{Sanket S. Kalamkar,
%\authorrefmark{1}\thanks {*This paper is a part of Master's thesis \cite{singh}, carried out at Department of Electrical Engineering, Indian Institute of Technology Kanpur, India.}, 
Praveen Kumar Singh, and Adrish Banerjee}
\IEEEauthorblockA{Department of Electrical Engineering, Indian Institute of Technology Kanpur, India}
\thanks{This work was supported in part by the TCS research scholarship program at Indian Institute of Technology Kanpur, India.
}
Email: \{kalamkar, adrish\}@iitk.ac.in}

\maketitle
\begin{abstract}
Block outlier detection methods, based on Tietjen-Moore (TM) and Shapiro-Wilk (SW) tests, are proposed to detect and suppress spectrum sensing data falsification (SSDF) attacks by malicious users in cooperative spectrum sensing. First, we consider basic and statistical SSDF attacks, where the malicious users attack independently. Then we propose a new SSDF attack, which involves cooperation among malicious users by masking. In practice, the number of malicious users is unknown. Thus, it is necessary to estimate the number of malicious users, which is found using clustering and largest gap method. However, we show using Monte Carlo simulations that, these methods fail to estimate the exact number of malicious users when they cooperate. To overcome this, we propose a modified largest gap method.
\end{abstract}
\vspace*{-5mm}
\section{Introduction} 
\subsection{Data Falsification in Cooperative Spectrum Sensing}
%Cooperative spectrum sensing (CSS) can be used to mitigate the harmful effects of multipath fading, shadowing and hidden terminal problem, faced by a single user spectrum sensing \cite{cabric}. In CSS, multiple secondary users (SUs) at different spatial locations, cooperate each other to effectively detect a primary user (PU), by exploiting the spatial diversity. In centralized approach, each SU reports local spectrum sensing data about PU to a fusion center (FC). Then FC takes the final decision on the presence or absence of PU, according to some fusion rule. However, the cooperation among SUs raises concerns about reliability and security of CSS, as some SUs may report falsified spectrum sensing data to FC. The falsified reported data can easily influence the spectrum sensing decision taken by FC. The falsification of data may occur either by malfunctioning of the SUs or by intentional manipulation of data by certain SUs, called malicious users (MUs). The data reported by malfunctioning SUs may differ from the actual data. In addition, the MUs can attack by manipulating the reported data with selfish intention, that is, to gain access to channel, or to cause interference to PU. Since in this attack, the spectrum sensing data is falsified, the attack is called spectrum sensing data falsification (SSDF) attack \cite{rchen}. 
In cooperative spectrum sensing (CSS), multiple secondary users (SUs) cooperate to effectively detect a primary user (PU), by exploiting the spatial diversity. However, the cooperation among SUs raises concerns about reliability and security of CSS, as some of the SUs may report the falsified spectrum sensing data to the fusion centre. The falsified reported data can easily influence the spectrum sensing decision taken by the fusion centre. The falsification of data may occur either by malfunctioning of SUs or by intentional manipulation of data by certain SUs, called malicious users (MUs). The data reported by malfunctioning SUs may differ from the actual data. In addition, MUs can attack by manipulating the reported data with selfish intention, i.e., to gain access to the channel, or to cause interference to PU. Since the spectrum sensing data is falsified in this attack, this is called spectrum sensing data falsification (SSDF) attack \cite{rchen}. 

An outlier is the data, that appears to be inconsistent with rest of the data \cite{Barnett1978}. The local spectrum sensing data reported by MUs may differ from the actual sensed data. Thus, MUs reporting the falsified spectrum sensing data, can be considered as outliers and detected using outlier detection techniques. \vspace*{-1mm}
\subsection{Summary of Results and Related Work}
\subsubsection{Summary of Results}
\begin{figure}
\centering
 \includegraphics[scale=0.28]{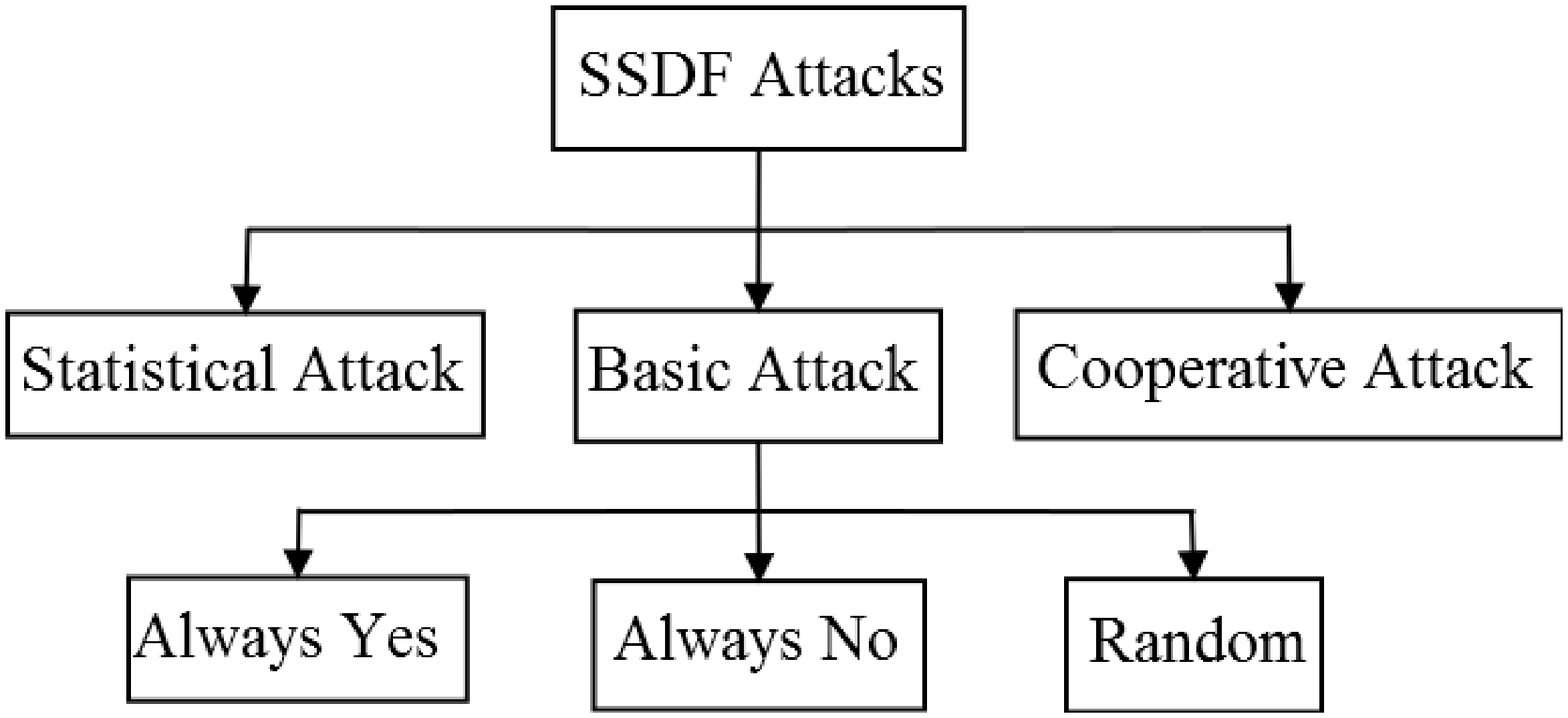}
 \caption{SSDF attack models}
 \label{fig:attack_model}
 \vspace*{-6mm}
 \end{figure}

Firstly, we propose two block outlier detection methods, based on Tietjen-Moore (TM) test \cite{Tietjen1972} and Shapiro-Wilk (SW) test \cite{Shapiro1965}, to counter different SSDF attacks in CSS as shown in Fig. \ref{fig:attack_model}, and compare them with box plot and median absolute deviation (MAD) tests \cite{le}. We show that TM and SW tests are more robust to SSDF attacks than the box plot and MAD tests. Secondly, we propose a new SSDF attack, called \textit{cooperative attack}, in CSS framework, which involves cooperation among MUs by masking. Thirdly, for cooperative attack, we propose a modified largest gap method, which can accurately estimate the number of outliers, required by TM and SW tests, whereas clustering \cite{macqueen} and the largest gap method \cite{Tietjen1972} fail to estimate the exact number of outliers.
\subsubsection{Related Work}
The basic SSDF attacks like ``Always Yes," ``Always No" and ``Random" are studied in \cite{rchen,FRYu2009}. In \cite{HLi2009, Penna2012}, the statistical attack is considered, where MUs act maliciously with a certain probability. However, in these attacks, no cooperation among MUs is considered. A consensus-based method is proposed to overcome the basic SSDF attacks \cite{FRYu2009}. To counter the statistical attack, in \cite{HLi2009}, an onion-peeling approach based on calculation of suspicious levels is adopted, while in \cite{Penna2012}, belief propagation is used. The reputation and weight based methods try to alleviate the detrimental effects of MUs by assigning trust values or weights to SUs based on the credibility of SUs, and are studied in \cite{TZhao2009, WWang2009, chen, zeng, arshad}. In \cite{duan}, a scenario is considered where multiple MUs can overhear the honest SU sensing data. Two attack-prevention mechanisms based on direct and indirect punishment are proposed to foil such attacks.

In \cite{kaligineedi2008}, an outlier detection method is proposed to pre-filter the extreme data, followed by the calculation of weights based on mean of the received spectrum sensing data. This method is further extended in \cite{kaligineedi2010}, where the outlier factors are calculated using weighted sample mean and standard deviation of the received sensing data. Based on the dynamic PU activity and the sensing data from the closest SUs, the outlier factors are adjusted. In \cite{le}, the detection performances of different outlier methods like MAD, box plot and median rule, are compared under low SNR nodes scenario (similar to ``Always No" attack). Anderson-Darling goodness-of-fit test is used in \cite{noh} to detect MUs by checking whether empirical distribution of SUs fit the expected distribution of a MU. In \cite{kalamkar}, the outlier tests like Dixon's test, box plot and Grubbs' test are studied to detect a single MU with basic attacks in CSS.
  
\begin {table}
\caption {Main Notations}
\centering
%\begin{center}
    \begin{tabular}{| l | l |}
    \hline
    \textbf{Notation}  & \textbf{Meaning} \\ \hline
    $H_1$ &  Hypothesis when the primary user is present.  \\ \hline
    $H_0 $ &  Hypothesis when the primary user is absent.  \\ \hline
    $ M$ &  Number of sensing samples.  \\ \hline   
    $ \alpha $  & Received signal-to-noise ratio (SNR) at SU.\\ \hline
    $ P_{FA} $ & Probability of false alarm at a single SU.\\ \hline
    $P_D$  & Probability of detection at a single SU. \\ \hline  
    $Q_{FA}$  & Probability of false alarm at fusion centre. \\ \hline  
    $Q_{D}$  & Probability of detection at fusion centre. \\ \hline 
    $L$  & Number of malicious SUs. \\ \hline 
    $P$  & Number of honest SUs. \\ \hline 
    $N$  & Number of cooperating SUs. \\ \hline 
     \end{tabular}
     \label{tab:simparam} \vspace*{-5mm} 
%\end{center}
\end{table}  
      
\section{System Model}
Consider a centralized CSS scenario with $N$ SUs, one PU and a fusion centre. Secondary users perform local spectrum sensing using energy detection \cite{urkowitz} and report the sensed energies to the fusion centre. Let $H_1$ and $H_0$ denote the binary hypothesis corresponding to the presence and absence of PU respectively. Then the detection problem of PU can be formulated as follows:\vspace*{-1mm} 
\begin{equation}
 y(m) = \left\{
  \begin{array}{l l}
    s(m) + u(m), & \quad H_1, \\
    u(m), & \quad H_0,\\
  \end{array} \right.
  \label{eq:binary}\vspace*{-1mm} 
\end{equation}
where $y(m)$ is the $m$th sample of the received signal by a SU with $m = 1, \dotsc, M$, $s(m)\sim \mathbb{CN}(0,\sigma_s^2)$, is the PU signal, and  $u(m)\sim \mathbb{CN}(0,\sigma_u^2)$, is additive white Gaussian noise (AWGN). We assume that $s(m)$ and $u(m)$ are independent. The primary signal samples are assumed to be independent. Also, we assume that the noise samples are independent. The received signal-to-noise ratio (SNR) is $\alpha=\frac{\sigma_s^2}{\sigma_u^2}$.

The test statistic $T$ for the energy detector is given by \cite{urkowitz}\vspace*{-1mm} 
\begin{eqnarray}
T(y)=\frac{1}{M}\sum_{m=1}^{M}{|y(m)|}^2,\vspace*{-1mm} 
\end{eqnarray}
and it is chi-squared distributed. However, for large $M$ ($M>$ 10), the test statistic can be approximated by Gaussian distribution according to central limit theorem. For this, the expressions of the probability of false alarm $P_{FA}$ and the probability of detection $P_D$ are given as follows \cite{liang}:\vspace*{-2mm} 

{{\small
\begin{eqnarray}
P_{FA}=Q\left[\left(\frac{\lambda}{\sigma_u^2}-1\right)\sqrt{M}\right]; P_{D}=Q\left[\left(\frac{\lambda}{\sigma_u^2}-\alpha-1\right)\frac{\sqrt{M}}{\alpha+1}\right], \nonumber
\end{eqnarray}}}
where $\lambda$ is a predetermined threshold. The fusion centre uses Majority logic \cite{liang} as a fusion rule, i.e., the final decision taken by the fusion centre is consistent with the local decisions taken by majority of SUs. The probability of false alarm and the probability of detection after fusing the local decisions at the fusion centre are denoted by $Q_{FA}$ and $Q_D$ respectively.
%
%{{\small 
%\begin{equation}
%Q_{FA} = \displaystyle \sum_{j=0}^{N-\lceil \frac{N}{2} \rceil} \binom {N} {\lceil \frac{N}{2}\rceil + j}\left(1-{P_{FA}}\right)^{N-\lceil \frac{N}{2} \rceil-j}{{P_{FA}}}^{\lceil \frac{N}{2} \rceil +j}
%\label{eq:majpf}
%\end{equation}}}
% and
%{{\small
%\begin{equation}
%Q_D = \displaystyle \sum_{j=0}^{N-\lceil \frac{N}{2} \rceil} \binom {N} {\lceil \frac{N}{2} \rceil + j}\left(1-{P_D}\right)^{N-\lceil \frac{N}{2} \rceil-j}{{P_D}}^{\lceil \frac{N}{2} \rceil +j}, 
%\label{eq:majpd}
%\end{equation}}}
%where $\binom{a}{b} = \dfrac{a!}{b!(a-b)!}$ and $\lceil x \rceil$ denotes the ceiling function.

Let $L$ and $P$ be the number of the malicious and the honest SUs respectively. We assume that the majority of SUs are honest ($L$ $<$ $P$). Thus, the falsified spectrum sensing data (falsified energy values in our case) by MUs, do not agree with the majority of the data reported by the honest users. Using outlier techniques, such malicious SUs are detected, and removed from cooperation to detect PU. The final decision about PU is made by fusing the local spectrum sensing decisions of only honest SUs.\vspace*{-1mm} 
\section{Cooperative Attack}
\label{sec:attack}

The basic and statistical attack models assume that MUs act independently, i.e., they do not cooperate among themselves. However, the more effective SSDF attacks may be launched if MUs cooperate with each other. In the proposed model, we consider that MUs cooperate using \textit{masking}. It is seen that the outlier tests suffer from the problem of masking \cite{bendre}. In masking, there exists extreme as well as not-so-extreme (mild) outliers. The extreme outliers modify the test statistic of an outlier test used to detect outliers such that, the presence of not-so-extreme outliers is shadowed by the extreme outliers, i.e., the outlier test fails to detect the not-so-extreme outliers, and only the extreme outliers are detected.

In the framework of CSS with energy detection, masking can be done as follows: A fraction of MUs report significantly different energy values than the actual sensed values, and the remaining MUs report slightly different energy values than the actual sensed values. This alters the test statistic used by an outlier test to detect outliers. The alteration in the test statistic is made such that, MUs which have reported slightly different energy values, escape from getting declared as outliers. Thus, they can continue to send the falsified  sensing data to the fusion centre to influence the spectrum sensing decision. \vspace*{-1mm} 

\section {Multiple Outliers Detection}
The multiple outliers can be present in three locations of the sorted data as follows:
\begin {itemize}
\item Upper outlier: Unexpected large values.
\item Lower outlier: Unexpected small values.
\item Bi-directional outliers: Both upper and lower outliers are present.
\end{itemize}

An outlier test should be able to identify all types of outliers. To know the type of an outlier, it is required to apply outlier tests designed for upper, lower or bi-directional outliers, on the received data. The data declared as outliers, are categorized as upper, lower or bi-directional outliers, based on which outlier test has detected them as outliers. An outlier test can be applied using either of the following two procedures:
\begin {itemize}
\item  Consecutive procedure: It is also called recursive procedure, that makes repeated use of a single outlier detection test, to remove outliers one by one. However, it is inappropriate to use a test for a single outlier detection recursively to detect multiple outliers \cite{Barnett1978}. Also, even though consecutive tests are easy to apply, they are inefficient for large data with many outliers.
\item  Block procedure: In this procedure, the outliers are tested in a block. The test requires calculation of the test statistic based on the received data. The test statistic is compared with the critical value, and based on this comparison, the whole block of data is adjudged as either outlier or non-outlier. In this paper, we consider two multiple outliers detection tests based on block procedure as follows:
\begin {itemize}
\item  Tietjen-Moore (TM) test
\item  Shapiro-Wilk (SW) test.
\end{itemize}
\end{itemize}
\subsection{Tietjen-Moore Test}
Tietjen and Moore proposed three test statistics \cite{Tietjen1972} to detect multiple outliers. All types of outliers, whether upper, lower or bi-directional, can be tested by choosing a suitable test statistic. The algorithm to detect upper or lower outliers is given by Algorithm \ref{algo:TMU}.\vspace*{-2mm} 
\begin{algorithm}
\caption{TM Test for Upper/Lower Outliers}
\begin {itemize}
\item[1:]  Sort the received energy values $y_1,\dotsc,y_N$ of $N$ SUs in ascending order. Let this sorted values be denoted by $x_1,\dotsc,x_N$.
\item[2:] Estimate $t$, the number of outliers (discussed in Section \ref{sec:est}).
\item[3:]  Calculate the test statistic given in \eqref{eq:TM_upper1} (for upper outliers), or given in \eqref{eq:TM_lower1} (for lower outliers).  
\item[4:]  Compare this test statistic with the critical value for significance level of 0.05, from the table given in \cite{Tietjen1972}. 
\item[5:]  If the test statistic is less than the critical value, then the suspected data are declared as outliers. 
\end{itemize}
\label{algo:TMU}
\end{algorithm}\vspace*{-2mm} 

The test statistic for testing upper outliers is as follows \cite{Tietjen1972}:  
\begin{eqnarray}
T = \frac{\sum_{j=1}^{N-t}(x_j-\bar{x}_t)^2}{\sum_{j=1}^{N}(x_j-\bar{x})^2},
\label{eq:TM_upper1}
\end{eqnarray}
where $\bar{x}$ is the sample mean and $\bar{x}_t$ is given by $\bar{x}_t = \frac{\sum_{j=1}^{N-t} x_j}{N-t}$.
%\begin{eqnarray}
%\bar{x}_t = \frac{\sum_{j=1}^{N-t} x_j}{N-t}.
%\label{eq:TM_upper2}
%\end{eqnarray}
Similarly, the test statistic to test lower outliers is given by \cite{Tietjen1972}
\begin{eqnarray}
T^*= \frac{\sum_{j=t+1}^{N}(x_j-{\bar{x}}_t^*)^2}{\sum_{j=1}^{N}(x_j-\bar{x})^2},
\label{eq:TM_lower1}
\end{eqnarray}
where ${\bar{x}}_t^* = \frac{\sum_{j=t+1}^{N} x_j}{N-t}$.
%\begin{eqnarray}
%{\bar{x}}_t^* = \frac{\sum_{j=t+1}^{N} x_j}{N-t}.
%\label{eq:TM_lower2}
%\end{eqnarray}
The algorithm for applying TM test to detect bi-directional outliers is as follows:\vspace*{-2mm} 
\begin{algorithm}
\caption{TM Test for Bi-directional outliers}
\begin {enumerate}
\item[1:]  Compute the mean $\bar{y}$ of the received energy values $y_1,\dotsc,y_N$.
\item[2:]  Compute $N$ absolute residuals $r_j =|y_j-\bar{y}|$, where $j=1,\dotsc,N$.
\item[3:]  Arrange $r_j$s in ascending order. Let this arranged data be denoted by $x_1,\dotsc,x_N$.
\item[4:] Estimate $t$, the number of outliers.
\item[5:] Calculate the test statistic as per \eqref{eq:TM_upper1}. 
\item[6:]  Perform steps 4 and 5 of Algorithm \ref{algo:TMU}.
\end{enumerate}
\label{algo:TMB}
\end{algorithm}\vspace*{-4mm} 

\subsection{Shapiro-Wilk Test}
This test, proposed by Shapiro and Wilk \cite{Shapiro1965}, is composed by considering a linear combination of ordered data, squaring it and then dividing it by an estimate of variance. The proposed test statistic is location and scale invariant, and is suitable for all types of data. The algorithm to apply SW test is given by Algorithm \ref{algo:SW}. The test statistic for SW test is given as \cite{Shapiro1965}
\begin{eqnarray}
T =\frac{\sum_{j=1}^{[\frac{N}{2}]}a_{N,N-j+1}(x_{N-j+1}-x_j)^2}{S^2},
\label{eq:SW_test}
\end{eqnarray}
where,\vspace*{-3mm} 
\begin{eqnarray}
S^2= \sum_{j=1}^{N}(x_j-\bar{x})^2 \hspace*{2mm} \text{with} \hspace*{2mm} \bar{x}=\frac{\sum_{j=1}^{N}x_j}{N}.
\label{eq:SW_test1}
\end{eqnarray}\vspace*{-3mm}

Here, $[\frac{N}{2}]$ is the integer part of $\frac{N}{2}$, and $a_{N,j}$ is the tabulated constant. The tables of tabulated constants and the critical values for significance level of 0.05 are given in  \cite{Shapiro1965}.\vspace*{-2mm}
\begin{algorithm}
\caption{SW Test}
\begin {itemize}
\item[1:] Sort the received energy values $y_1,\dotsc,y_N$ of $N$ SUs in ascending order. Let this sorted values be denoted by $x_1,\dotsc,x_N$.
\item[2:] Estimate $t$, the number of outliers.
\item[3:] Calculate the test statistic as per \eqref{eq:SW_test} and \eqref{eq:SW_test1}.
\item[4:]  If the test statistic is less than the critical value for significance level of 0.05, then the suspected data are declared as outliers.
\end {itemize}
\label{algo:SW}
\end{algorithm}\vspace*{-2mm}

%
%\subsection{Box Plot Test}
%Though not a block outlier test, the box plot method can be used to detect multiple outliers. In box plot method \cite{boxplot}, the received energy values are arranged in ascending order, and the SU is considered malicious, if the received energy of SU falls below  $Q_L$ (lower outlier) or above $Q_U$ (upper outlier). The threshold values $Q_L$ and $Q_U$ are calculated as follows:
%\begin{equation}
%Q_{L} = Q_1 - 1.5 Q_{int}, \hspace*{2mm} \text{and} \hspace*{2mm} Q_{U} = Q_3 + 1.5 Q_{int},
%\end{equation}
%where $Q_1$ is first quartile, $Q_3$ is third quartile and $Q_{int}$ is $Q_3 - Q_1$, that is, the interquartile range.
%\subsection{Median Absolute Deviation (MAD) Test}
%MAD test also can be used to detect multiple outliers. In MAD \cite{tukey}, the received energy values are arranged in ascending order as $x_1,\dotsc,x_N$, and the lower $Q_L$ and the upper $Q_U$ threshold values, to detect outliers, are calculated as follows: 
%\begin{equation}
%Q_{L} = Q_2 - 3 MAD_e, \hspace*{2mm} \text{and} \hspace*{2mm} Q_{U} = Q_2 + 3 MAD_e,
%\end{equation}
%where $Q_2$ is second quartile, $MAD_e = 1.483 \times MAD$. $MAD$ is median absolute deviation, defined as, $MAD = \text{median}|x_i - Q_2|_{i = 1,\dotsc,N}$. The value 1.483 is the constant scale factor for Gaussian distribution.

\subsection{Estimating the Number of Outliers} 
\label{sec:est}
In practice, the number of MUs is not known. Also, to apply TM and SW tests, an estimate of the number of outliers is required. For this, we consider clustering and the largest gap method proposed in the literature, and are described as follows:

\subsubsection{Clustering method} We consider clustering as a tool to estimate the number of outliers. We use $k$-means clustering algorithm \cite{macqueen} to group the data set into two clusters. The smaller of these clusters is treated as a cluster of suspected outliers, and is tested against TM and SW tests, to decide whether this assumption is true or not.  
\subsubsection{Largest gap method} Tietjen and Moore proposed a method based on the largest gap between the data points, as a basis to estimate the number of outliers \cite{Tietjen1972}. For upper/lower outliers, the procedure for the largest gap method is as follows:\vspace*{-2mm}
\begin{algorithm}
\caption{Largest Gap method for Upper/Lower Outliers}
\begin{itemize}
\item[1:] Sort the received data (energy values) in ascending order for upper outliers (descending order for lower outliers).
\item[2:] Calculate the gaps between successive data points.
\item[3:] Find the position of the largest gap.
\item[4:] The number of data points to the right of this position gives an estimate of number of outliers.
\end{itemize}
\label{algo:ul}
\end{algorithm}

The largest gap method for bi-directional outliers is given as follows:
\begin{algorithm}
\caption{Largest Gap Method for Bi-directional Outliers}
\begin{itemize}
\item[1:]	Sort the received data set (energy values) $D$ in ascending order and divide it into two halves, lower half $D_{LH}$ and upper half $D_{RH}$.
\item[2:]	Apply largest gap method for upper outliers, proposed in Algorithm \ref{algo:ul} to $D_{RH}$ and  largest gap method for lower outliers to $D_{LH}$, to get an estimate of number of outliers in both halves.
\item[3:]  Apply TM/SW test on both $D_{RH}$ and $D_{LH}$ separately, to decide which half contains outliers.
\end{itemize}
\label{algo:bil}
\end{algorithm}

We show in Section \ref{results} that, when MUs launch cooperative attack using masking, both clustering and largest gap method fail to estimate the correct number of outliers. Thus, to overcome this, we propose a modified largest gap (MLG) method, which can give the suspected number of outliers accurately under cooperative malicious users attack.

\subsection{Proposed Method: Modified Largest Gap}
The proposed modified largest gap method involves applying the largest gap method recursively until all the outliers are detected. The algorithm to apply MLG for detection of upper outliers is given as follows:
\begin{algorithm}
\caption{MLG method for Upper Outliers}
\begin{itemize}
\item[1:]   Sort the received data set (energy values) $D'$ in ascending order to obtain $D$. 
\item[2:] 	Divide D in two halves: Lower half ($D_{LH}$) and upper half ($D_{UH}$).
\item[3:] 	As the number of outliers is in minority, $D_{UH}$ consists of the energies reported by malicious users.
\item[4:] 	Calculate the gap between successive data points in $D_{UH}$.
\item[5:] 	Find the position of the largest gap (denoted by ${G}_{pos}$).
\item[6:] 	Let $S_1$ be the set of data points to the left of the ${G}_{pos}$ in $D_{UH}$, and let $S_2$ be the set of data points to the right of the ${G}_{pos}$ in $D_{UH}$. Form a new set $S = D_{LH} \cup S_1 $. 
\item[7:] 	Test whether $S_2$ is outlier (using TM test).
\item[8:] 	If $S_2$ is outlier, then reject $S_2$; let $D=S$, and go to Step 2. Else, $S \cup S_2$ is a data set of honestly reported energy values.
\end{itemize}
\label{algo:up}
\end{algorithm}

For lower outliers, the MLG method is the same as that of upper outliers, except that the data is sorted in descending order. The procedure to apply MLG for bi-directional outliers is same as Algorithm \ref{algo:bil}, except the largest gap method is replaced by the MLG method.

\section{Simulation Results}
In this section, we present Monte Carlo simulation results to show the spectrum sensing performance of CSS with outlier tests, to defend different SSDF attacks as shown in Fig. \ref{fig:attack_model}. In simulations, we have considered the following parameters: $\sigma_u^2$ = 1, $M$ = 10000, $N$ = 20, $L$ = 4. The energy value reported by a MU differs from the actual sensed value by 0.5 dB.
\label{results} 
\subsection {Suppressing malicious users in the basic attack model} 
\begin{figure}
        \centering
        \includegraphics[scale=0.43]{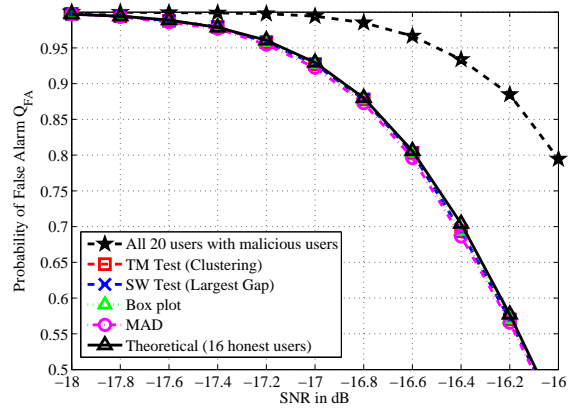}
        \caption{``Always Yes" attack: Comparison of outlier tests, $Q_D$ = 0.99.}
        \label{fig:yes_cot}
        \end{figure}

\begin{figure}
        \centering
        \includegraphics[scale=0.43]{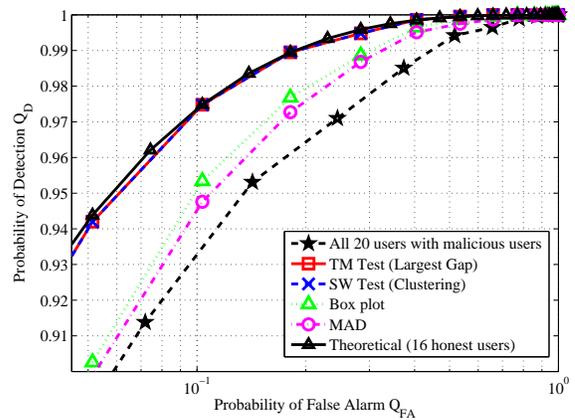}
        \caption{``Random" attack: Comparison of outlier tests, SNR = -20 dB.}
        \label{fig:random_cot}
        \end{figure}

For ``Always Yes" attack (Fig. \ref{fig:yes_cot}), all the outlier tests considered in this paper are able to detect all the MUs successfully. It is shown that $Q_{FA}$ can be decreased by refraining MUs from participating in CSS, compared to when all SUs including MUs participate in CSS. Also, the outlier tests perform similarly for ``Always No" attack (results are not shown due to space constraint). However, for ``Random" attack (Fig. \ref{fig:random_cot}), TM and SW tests are more robust than the box plot and MAD tests. This is because TM and SW tests are able to detect all randomly behaving 4 MUs, but the box plot and MAD tests can detect only a fraction of MUs, giving worse performance. Both clustering and the largest gap method perform the same, when they are used to estimate the number of outliers (Figs. \ref{fig:yes_cot} and \ref{fig:random_cot}). It can also be noticed that TM and SW tests perform the same.
\subsection{Suppressing malicious users in statistical attack model} 
\begin{figure}
      \centering
      \includegraphics[scale=0.43]{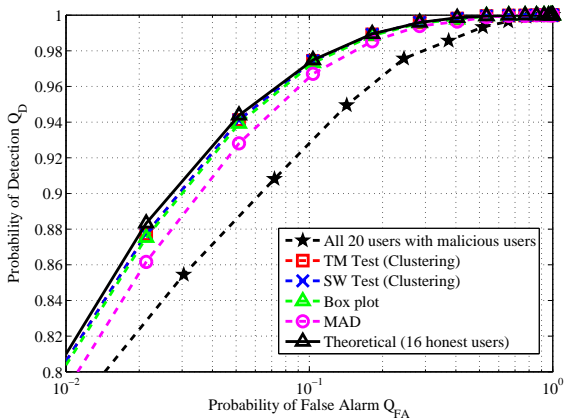}
      \caption{Statistical attack: Comparison of outlier tests, SNR = -20 dB.}
      \label{fig:stat}
      \end{figure}

It is shown in Fig. \ref{fig:stat} that, TM and SW tests are good enough to counter the statistical attack, when MUs act maliciously with a certain probability. Clustering is used to estimate the number of MUs for both TM and SW tests. Box plot's detection performance is almost the same as TM and SW tests, while MAD test performs the worst.  
                 
\subsection{Suppressing malicious users in cooperative attack model}
\begin{figure}
\centering
\includegraphics[scale=0.43]{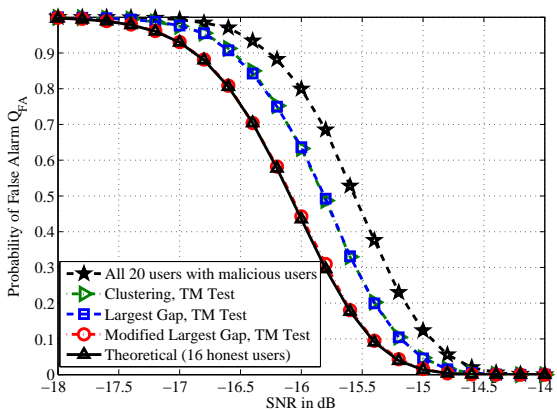}
\caption{Cooperative ``Always Yes" attack: TM test with different methods to estimate number of outliers, $Q_D$ = 0.99.}
\label{fig:coop} \vspace*{-1mm}  
\end{figure}

As aforementioned, MUs may cooperate using masking. We consider that all MUs are upper outliers. Masking is done, as half of the outliers are extreme outliers reporting significantly high energy values (6.5 dB greater than the actual sensed energy in our case), and the rest are mild outliers reporting slightly higher energy values than the actual energy values (0.5 dB greater than the actual sensed value). Then, as shown in Fig. \ref{fig:coop}, TM or SW tests fail to counter cooperative attack when clustering or the largest gap method is used to estimate the number of MUs. In clustering, the smaller cluster might consist of only extreme outliers, and no mild outliers as the latter ones may be masked by the former ones. For the largest gap method, the largest gap occurs between mild outliers and extreme outliers, as the energies reported by the extreme outliers are significantly different from rest of the energy values. However, the proposed modified largest gap method is highly effective in estimating the accurate number of cooperating MUs, as it finds the largest gap recursively until all the outliers are detected, thus nullifying their harmful effects.
\section{Conclusions}
In this paper, two block outlier tests, TM test and SW test, are proposed to suppress an unknown number of the malicious users in cooperative spectrum sensing, and compared with box plot and MAD tests. We have shown that TM and SW tests are more robust than the box plot and MAD tests for ``Random" and statistical attacks. We have also proposed a cooperative SSDF attack, which adopts cooperation among malicious users by masking, where the presence of extreme outliers mask the mild outliers. Also, it is shown using Monte Carlo simulations that, clustering and the largest gap method fail to accurately estimate the number of outliers in cooperative attack. Thus, to overcome this shortcoming, we propose a modified largest gap method, which can accurately estimate the number of outliers under cooperative attack.
\bibliography{thesis_ref}
\bibliographystyle{ieeetr}
\end{document}